\documentstyle[emulateapj5]{aastex}

\def\simgt{\lower.5ex\hbox{$\; \buildrel > \over \sim \;$}}
\def\simlt{\lower.5ex\hbox{$\; \buildrel < \over \sim \;$}}
\def\sp{\hspace{1.5pt}}

\def\axp{\hbox{AX~J1844.8$-$0256}}
\def\taxp{\hbox{XTE~J1810$-$197}}
\def\src{\hbox{CXOU~J180951.0$-$194351}}

\def\s{\phantom}
\def\ss{\phantom\phantom}

\def\ssss{\phantom\phantom\phantom\phantom}
\def\sp{\hspace{1.5pt}}

\def\amin{\ifmmode^{\prime}\else$^{\prime}$\fi}
\def\asec{\ifmmode^{\prime\prime}\else$^{\prime\prime}$\fi}

\def\simgt{\lower.5ex\hbox{$\; \buildrel > \over \sim \;$}}
\def\simlt{\lower.5ex\hbox{$\; \buildrel < \over \sim \;$}}

\newcommand\xte{{\it RXTE\/}}

\newcommand\einstein{{\it Einstein}}

\newcommand\asca{{\it ASCA\/}}

\newcommand\rosat{{\it ROSAT\/}}
\newcommand\chandra{{\it Chandra}}
\newcommand\Chandra{{\it Chandra}}
\newcommand\xmm{{\it XMM\/}-Newton}

\def\sp{\hskip 1.5pt}

\def\kes73{\hbox{Kes\sp73}}

\shorttitle{Anomalous X-ray Pulsar XTE J1810--197}
\shortauthors{Gotthelf et al.}

\received{2003 September 26}
\begin{document}

\title{Imaging X-ray, Optical, and Infrared Observations of the Transient Anomalous X-ray Pulsar XTE J1810--197}
\author{E. V. Gotthelf$^1$, J. P. Halpern$^1$, M. Buxton$^2$, C. Bailyn$^2$}
\altaffiltext{1}{Columbia Astrophysics Laboratory, Columbia University, 550 West
120$^{th}$ Street, New York, NY 10027, USA; evg@astro.columbia.edu; jules@astro.columbia.edu}
\altaffiltext{2}{Department of Astronomy, Yale University, P.O. Box
208101, New Haven, CT, 06520-8101, USA; bailyn@astro.yale.edu; buxton@astro.yale.edu}

\begin{abstract}

We report X-ray imaging, timing, and spectral studies of \taxp, a
5.54~s pulsar discovered by Ibrahim et al. (2003) in recent \xte\
observations. In a set of short exposures with the High Resolution
Camera on-board the \Chandra\ X-ray Observatory we detect a strongly
modulated signal ($55 \pm 4\%$ pulsed fraction) with the expected
period located at (J2000) $18^{\rm h}09^{\rm m}51^{\rm s}\!.08$,
$-19^{\circ}43^{\prime}51\farcs7$, with a uncertainty radius of
$0\farcs6$ (90\% confidence level). Spectra obtained with the Newton
X-Ray Multi-Mirror Mission (\xmm) are well fitted by a two-component
model that typically describes anomalous X-ray pulsars (AXPs), an
absorbed blackbody plus power law with parameters $kT = 0.67\pm 0.01$
keV, $\Gamma = 3.7\pm 0.2$, $N_{\rm H} = (1.05\pm 0.05) \times
10^{22}\ {\rm cm}^{-2}$, and $F_X(0.5-10\,{\rm keV}) = 3.98 \times
10^{-11}$ ergs cm$^{-2}$ s$^{-1}$. Alternatively, a two-temperature
blackbody fit is just as acceptable. The location of \src\ is
consistent with a point source seen in archival \einstein, \rosat, and
\asca\ images, when its flux was nearly two orders-of-magnitude
fainter, and from which no pulsations are found.  The spectrum changed
dramatically between the ``quiescent'' and ``active'' states, the
former can be modeled as a softer blackbody.  Using \xmm\ timing data,
we place an upper limit of 0.03 lt-s on any orbital motion in the
period range 10~min -- 8~hr. Optical and infrared images obtained on
the SMARTS 1.3~m telescope at CTIO show no object in the \Chandra\
error circle to limits $V = 22.5, I = 21.3, J = 18.9$, and $K = 17.5$.
Together, these results argue that \src\ is an isolated neutron star,
one most similar to the transient AXP \axp. Continuing study of \taxp\
in various states of luminosity is important for understanding and
possibly unifying a growing class of isolated, young neutron stars
that are not powered by rotation.
\end{abstract}


\section{Introduction}

X-ray observations over the past two decades have discovered young,
isolated neutron stars (NSs) with spectral and timing properties markedly
different from those of the typical rotation-powered (or
accretion-powered) pulsar. These objects, which tend to be labeled by
their most distinctive characteristic, include the anomalous X-ray
pulsars (AXPs) with slow ($5-12$~s) rotation periods and X-ray
emission far in excess of their rotational energy-loss rate, the Soft
Gamma-ray Repeaters (SGRs), sources of random episodes of short ($\sim
1$~s) bursts of hard X-rays which are otherwise similar to the AXPs,
and several examples of so-called Central Compact Objects (CCOs)
within supernova remnants (SNRs) that lack detectable periodicity but
have spectra resembling those of the AXPs and SGRs [e.g., the X-ray
point source in the center of Cas~A; Chakrabarty et al. (2001);
Mereghetti et al. (2002)]. Collectively, this population of uniquely
X-ray discovered objects likely account for a significant fraction of
young neutron stars (Gotthelf \& Vasisht 2000).

These Galactic plane residents have in common a notably soft X-ray
spectrum, a lack of radio or optical counterpart, and luminosities
distinct from most rotation-powered pulsars.  These properties, along
with a lack of detected orbital motion, no red noise in their light
curves, and a large X-ray to infrared flux ratio, also argue against
an accretion-driven origin.  Instead, these objects are understood in
terms of the magnetar model for the SGRs, in which the X-ray emission
is derived from magnetic energy loss in an strongly magnetized ($B >
4.4 \times 10^{13}$ G) isolated neutron star (Duncan \& Thompson 1992).  Most
mysterious then was the discovery of a transient AXP (TAXP), \axp\
(Gotthelf \& Vasisht 1998; Torii et al. 1998). This 7-s pulsar was
seen just once in archival \asca\ data, and has since faded in flux by
at least an order of magnitude.  It was not detected in earlier data
sets or {\it post facto} in an ongoing monitoring campaign.  But
otherwise, \axp\ has spectral and timing properties that are most
consistent with an AXP.

The newly discovered X-ray pulsar \taxp\ (Ibrahim et al. 2003) is likely
a second example of a TAXP. Archival \xte\ data show that \taxp\
became active in 2003 January; as of 2003 August its flux has
faded by half (Markwardt et al. 2003a,b).  Long-term timing of \taxp\
with \xte\ suggests an AXP based on the period
derivative and lack of detected orbital motion (Ibrahim et al. 2003).

In this paper, we report results of X-ray, optical, and infrared
observations of \taxp\ that enabled us to locate the pulsar with
arcsecond precision, measure its X-ray spectrum, and search for an
optical or infrared counterpart.  These prompt follow-up observations
comprise a unique program to study the evolution of a transient AXP
while it is bright and active as a pulsar.  Based on these
observations, we propose that TAXPs, AXPs, SGRs, and CCOs are
inherently a single class of young objects observed in various stages
of activity but whose emission physics
is distinct from rotation-powered pulsars.

\section{Chandra Observations and Results}

Based on the report of a new pulsar by Markwardt, Ibrahim, \& Swank
(2003a), we submitted a Target of Opportunity request to the \Chandra\
X-ray Observatory (Weisskopf, O'Dell, \& van Speybroeck 1996) to
localize \taxp\ while the source of the pulsations was still
active. The observation was carried out on 2003 August 27 using the
\Chandra\ imaging High Resolution Camera (HRC-I; Murray et al. 1997),
a multichannel plate detector sensitive to X-rays over the $0.08-10.0$
keV range and with a timing precision of $<4$~ms, depending on detector
count rate, more than sufficient to resolve the 5.54~s pulsations. No
energy information is recorded for photons detected with the HRC. The
observation was centered on the reported $11^{\prime} \times
20^{\prime}$ diameter \xte\ 99.73\% confidence error ellipse, with the roll angle
adjusted to avoid bright diffuse emission from the nearby supernova
remnant G11.2--0.3 and its central pulsar (see Figure~1).

The HRC data collected during the observation shows long periods of
high background activity, most likely of solar wind origin, with a
mean background rate over the detector of $70.5$ counts~s$^{-1}$.
Although telemetry saturation occurred during 65\% of the observation,
this does not create any additional errors in the detected event times
(M. Juda, personal communication).  The standard Level~2 HRC
processing produced a total good exposure time for the observations of
2.8~ks.

Figure~2 displays the HRC-I image rebinned into $2\farcs1 \times
2\farcs1$ pixels and smoothed with a Gaussian of $\sigma =
3\farcs2$. Only one bright source is apparent, $14^{\prime}\!.8$
off-axis, with a background-subtracted count rate of $0.96 \pm 0.04$
s$^{-1}$ after correcting for telemetry saturation.  The radial
profile of this source is consistent with an unresolved point source
at the off-axis HRC-I position, where the HRC point-spread function
(PSF) has a half-power diameter (the radius enclosing $50\%$ of total
source counts) of $\approx 13^{\prime\prime}$ at $1.5$ keV, increasing
with energy. 

The centroid of this source, based on the standard processing, is
(J2000) $18^{\rm h}09^{\rm m}51^{\rm s}\!.13$,
$-19^{\circ}43^{\prime}51\farcs7$ with an uncertainty radius of
$1\farcs7$.  The positional uncertainty is dominated by the systematic
distortion of the PSF at the off-axis position ($1\farcs6$: e.g.,
Flaccomio et al. 2003), but it also includes random error associated
with the aspect reconstruction ($0\farcs6$; 90\% confidence level) and
formal measurement error ($\sigma/\sqrt N = 0\farcs3$) based on the
number of source photons $N$ and the standard deviation $\sigma$ of
the PSF.  No other X-ray source is visible in the image, in
particular, at the positions of \rosat\ sources 1--3 shown in
Figure~1, and thus no sufficiently bright fiducial X-ray source is
available in the HRC-I field of view to refine the astrometry. We
report the sole HRC source as \src\ and a search of the high-energy
archives shows that its coordinates are consistent with the \rosat\
all-sky survey source 1RXS 180951.5--194345 (Voges et al. 1999) and
the \asca\ source AX~J180951--1943 (Bamba et al. 2003). These
coordinates are also consistent with the location of an uncatalogued
\einstein\ IPC source, $30^{\prime}$ off-axis in a 1980 March 31
observation of SNR G11.2--0.3 (Harris et al. 1990). A complete X-ray
observation log is presented in Table~1.

We examined the light curve of \src\ for a pulsed signal that might
identify it as \taxp. A total of 2605 photons were extracted from a
$0\farcm5$ radius aperture centered on this bright source; the arrival
times of the photons were corrected to the solar system barycenter in
Barycentric Dynamical Time (TDB) using the JPL DE200 ephemeris
(Standish et al. 1992). The extracted photons included 252 background
events, estimated from counts in a surrounding annulus.  We first
performed a Fast Fourier Transform (FFT) on the extracted photons and
found a single, highly significant signal at 5.54s.  No evidence was
found in the FFT of red noise typical of an X-ray binary system.  We
then generated a periodogram using the $Z^2_1$ statistic (Rayleigh
test) on a range of test frequencies centered on the FFT detected
signal.  We find a highly modulated sinusoidal pulse with a period $P
= 5.5392 \pm 0.0008$~s (95\% confidence) with a background-subtracted
pulsed fraction ($f_{\rm p} \equiv N_{\rm pulsed} / N_{\rm total}$) of
$56 \pm 7\%$ (see Figure~3). The Rayleigh statistic for this signal is
$Z^2_1 = 338$, a certain detection. Table~2 summaries the timing
results presented herein, which  confirms the identification
of \src\ as \taxp\ (Gotthelf et al. 2003).

A second, on-axis, \chandra\ HRC observation centered on the pulsar
location presented above was performed on 2003 November 1 (Israel
et al. 2004, in preparation). The pulse profile, pulsed fraction, and
coordinates derived from this observation are consistent with the
previous HRC and XMM measurements.  The on-axis location of the short
exposure ensures a background free detection of the pulsar with a
measured pulsed fraction of $54 \pm 5\%$. The updated coordinates for
the unresolved ($\sim 1\farcs0$) point source are (J2000) $18^{\rm
h}09^{\rm m}51^{\rm s}\!.08$, $-19^{\circ}43^{\prime}51\farcs7$, with
a reduced position uncertainty radius of $0\farcs6$ (90\% confidence
level) assuming the standard \chandra\ on-axis uncertainty and
negligible measurement error. These values are consistent with those
reported by Israel et al. (2003).

\section{\xmm\ Observation and Results}

Having located the X-ray source associated with the pulsations of
\taxp, a Target of Opportunity observation was performed with the
\xmm\ X-ray Observatory (Jansen et al. 2001) to determine the spectral
properties of the source in the active state. This observation was
carried out on 2003 September 8. Here we concentrate on data obtained
with the European Photon Imaging Camera (EPIC, Turner et al. 2001).
EPIC consists of three CCD cameras, the EPIC-pn and the two EPIC-MOS
imagers.  One of the EPIC-MOS cameras was operated in {\tt
FullFrameMode} mode, which is read out in 2.7~s integrations over the
$30^{\prime}$ diameter field-of-view. Data from the second EPIC-MOS
camera data was obtained in {\tt SmallWindowMode} for which the FOV of
the central CCD was reduced to $2\farcm0 \times 2\farcm0$ but read out
with a shortened integration time of 0.3~s. The EPIC-pn was operated
in {\tt SmallWindowMode} mode, which provides 6~ms time resolution
over a limited field-of-view of $4\farcm3$ square. The EPIC
instruments are sensitive to X-rays in $0.1-10$ keV nominal range.

The initial observation of \taxp\ was lost due to a hardware
malfunction and the target was re-observed in two segments with a short
interruption in between.  We processed the Observation Data Files
(ODF) in-house using the standard \xmm\ Science Analysis System (SAS)
processing chains for each instrument (release version {\tt
20020605\_1701-5.3.3}), and the photon event list was further filtered
using the standard SAS criteria. We selected only CCD {\tt PATTERN}
$\le 4$ data for our EPIC-pn spectral analysis. A total of $\approx
11.5$ ks of good exposure time was acquired on the target with only
minor background contamination that was not significant enough to
warrant filtering out.

We find one bright source in the combined EPIC FOV whose coordinates,
based on the standard processing lies $0\farcs8$ from \src\ (see
Table~1), well within the quoted $4\farcs0$ (90\% confidence level)
uncertainty radius for XMM. The measured pulse period from the
combined EPIC-pn and MOS1 data is $5.539344 \pm 0.000019$~s
(Table~2). The error is the 95\% confidence level determined from the
$Z_1^2$ test. Pulse profiles from the EPIC-pn and MOS1 instruments
individually are shown in Figure~4.  The pulse peak is somewhat
narrower than a sinusoid, and the pulsed fractions increase smoothly
with energy from 36\% at less than 1~keV to $\approx 55\%$ above
5~keV.  Based on the errors in the period measurements, it is not
quite possible to phase connect the \chandra\ and \xmm\ observations
taken 12 days apart and thereby improve the period accuracy; the
extrapolated phase is uncertain by 0.6 cycles.

To search for a faint supernova remnant surrounding the pulsar, we
made a more restrictive filtering of the data to maximize the
sensitivity to diffuse emission. No enhancement of the diffuse flux is
found around the EPIC PSF, however we cannot rule out a small nebula
obscured by the pulsar emission (see Figure 5).The \chandra\ HRC
observation was too short to image a faint SNR.

We fit the pulsar's spectrum using data from the EPIC-pn CCD. The fast
read out of this instrument ensures that its spectrum is not affected
by photon pile-up. A $90,462$ photon source spectrum was accumulated
from a 1\farcm5 diameter aperture which enclosed $\sim 85\%$ of the
PSF. This spectrum was grouped into bins containing a minimum of 400
counts and fitted using the {\tt xspec} package. The spectrum cannot
be fitted with any single component, but it is well fitted by a
two-component spectral model consisting of a blackbody plus a power
law, which is typical of AXPs (e.g., Marsden \& White 2001).  Figure~6
shows the spectrum fitted with temperature $kT = 0.67\pm 0.01$ keV and
photon index of $\Gamma = 3.7\pm0.2$, with a fit statistic of
$\chi^2_{\nu} = 1.0$ for 193 degrees of freedom (DoF).  The column
density for this fit is $N_{\rm H} = (1.05 \pm 0.05) \times 10^{22}\
{\rm cm}^{-2}$ resulting in a $0.5-10$ keV absorbed flux of $3.98
\times 10^{-11}$ ergs~cm$^{-2}$~s$^{-1}$ , and an intrinsic
(unabsorbed) flux of $1.38 \times 10^{-10}$ ergs~cm$^{-2}$~s$^{-1}$
(spectral fits are summarized in Table~3).

We also found that a fit to a two-component blackbody plus blackbody
model was equally acceptable but required a lower column density of
$N_{\rm H} = (0.63 \pm 0.05) \times 10^{22}\ {\rm cm}^{-2}$.  In this
fit the power-law component of the previous model is replaced with a
cooler blackbody with $kT = 0.29\pm 0.03$ keV while the hotter
component remained nearly unchanged with a temperature of $kT =
0.70\pm 0.02$ keV. The flux for each of these components is given in
Table~3.

A preliminary analysis of the Reflecting Grating Spectrometer (RGS)
spectrum ($0.8-1.7$ keV) containing 3700 background subtracted counts
reveals no narrow spectral features with equivalent width of $>20$ eV,
the presence of which might be associated with a NS atmosphere in a
high magnetic field (A. Rasmussen, personal communication).

\section{Interpreting Previous X-ray Observations}

We can now assume the \xmm\ spectral results to derive the flux during
$\chandra$ HRC observations.  The background-subtracted count rate for
the pulsar in the HRC, corrected for dead-time and mirror vignetting,
is 1.32 s$^{-1}$ and 1.13 s$^{-1}$ for the first and second
observations, respectively. Assuming the shape of the \xmm\ spectrum
given above, the respective $0.5-10$~keV HRC flux is then $3.92 \times
10^{-11}$ ergs~cm$^{-2}$~s$^{-1}$ and $3.35 \times 10^{-11}$
ergs~cm$^{-2}$~s$^{-1}$, consistent with the \xmm\ derived flux and
records a $15\%$ decline in intensity between the HRC observations
separated by just over three months.

We next measured the flux from \taxp\ in the ``quiescent'' state using
the archival \einstein, \rosat, and \asca\ detections. The \rosat\
all-sky survey source 1RXS~180951.5--194345 was serendipitously
detected off-axis in four pointed \rosat\ PSPC observations spanning
1991-1993 (see table 3). In particular, the best spectrum was obtained
from the 1993 Apr 3 observation of SNR~G11.2--0.3 in which \taxp\ fell
$29^{\prime}$ off-axis. A total of 294 counts were collected within a
$2\farcm4$ source extraction aperture during the 5340~s of good
exposure time. These counts are sufficient to show that the spectrum
has changed significantly in the active state, as the \xmm\ model
produced an unacceptable fit to the \rosat\ data. Keeping the column
density fixed to the \xmm\ derived value but leaving the black-body
and power-law normalization constants free to be fitted independently
also resulted in an unacceptable fit. Instead, the best fit is
obtained with a simple blackbody of temperature of $0.18\pm0.02$ keV
($\chi^2_{\nu} = 0.8$ for 14 degrees of freedom), which is preferable
over a simple power-law model which, although resulting in a similar
fit statistic, required an index of $\Gamma = 6.0$. The absorbed flux
is then $F_X(0.5-10.0\ \rm{keV}) = 5.5 \times
10^{-13}$~ergs~cm$^{-2}$~s$^{-1}$, almost two orders-of-magnitudes
less than that recorded in the active state. Table 3 summaries our
flux measurement using either the above model or a direct fit to the
\rosat\ archival data.

With the above quiescent state spectral model we can estimate the
detectability of the pulsed signal of \src\ in the 1993 Apr 3 PSPC
data. Based on the pulse profile of \src\ and an estimate of the local
background in the PSPC aperture, we expect the \rosat\ source to have
a pulsed fraction of 39.7\% if most of the pulsed emission is within
the \rosat\ energy band. This should produce a $Z^2_1$ statistic of 32
for a sine wave, corresponding to a false detection probability of
$1.1 \times 10^{-6}$ per trial.  However, no significant signal was
detected within a range encompassing any likely spin-down rate
($\pm0.05$~s) around the expected period to an upper limit of $Z^2_1 =
12$, corresponding to pulsed fraction $<24\%$.  We therefore
conclude that 1RXS 180951.5--194345 was less pulsed in the \rosat\
observation than it is in the current high state, otherwise we would
have easily detected the 5.54~s signal.

AX J180951--1943 was observed in three \asca\ observations, acquired
on 1996 April 2 and 8, and 1999 Sep 28.  Having data from the same
instrument should provide a measure of the variability, in
principle. However, these detections are at the $6\sigma$, $3\sigma$,
and $5\sigma$ significance level, respectively, as the location of the
source in each case fell partially off the edge of the GIS detector.
Notwithstanding, we find that the \rosat\ model gives a reasonable fit
to \asca\ data and no significant variation in flux is found between
these measurements taken 3.5 years apart (see Table~3). However, we
advance these \asca\ results with caution as the flux measurements are likely
affected by large unknown systematic uncertainties. Although no
periodic signal is detected in the \asca\ data, the search is hardly
constraining because of the low signal-to-noise ratio.

\section{Search for an Optical/IR Counterpart}

The identification of an optical/IR counterpart, or the apparent lack
of one, is key to interpreting the nature of \taxp.  A search of
stellar catalogs reveals no candidate within the HRC error circles 
of \src, nor does an inspection of the Digitized Sky Survey
(DSS) $B$-, $R$- and $I$-band images from either the first or second
generation plates. The closest star is $4\farcs7$ from the X-ray
position.  Listed as USNO-B1.0 ID\#0702-0541769 at $18^{\rm h}09^{\rm
m}51^{\rm s}\!.337$, $-19^{\circ}43^{\prime}48\farcs69$ (J2000), its
magnitudes are $R=18.72$ and $I=17.18$.  This object is also a 2MASS
source with measured infrared magnitudes of $J=14.744\pm0.028$,
$H=13.977\pm0.023$, and $K=13.854\pm0.056$.

We performed a deeper search for a stellar counterpart shortly
following the \chandra\ localization of \src\ using the
SMARTS\footnote{http://phoenix.astro.yale.edu/smarts/} 1.3m telescope
at CTIO on UT 2003 August 31.08 and September 01.14. Images in $V$-,
$I$-, $J$-, and $K$-band filters were obtained with the
ANDICAM\footnote{http://www.astronomy.ohio-state.edu/ANDICAM/}, a
dual-channel imager capable of obtaining optical and infrared data
simultaneously.  Data was recorded by a Fairchild-447 $2048\times2048$
CCD on the optical channel and a Rockwell $1024\times1024$ HgCdTe
``Hawaii'' array on the infrared channel.

Observations on both nights were taken through cirrus clouds. The
exposure times for the optical frames were 300~s, and a total of 210~s
for the infrared exposures. The final infrared images consisted of
seven 30~s sub-exposures, each shifted $\approx 30^{\prime\prime}$ in
right ascension or declination by an internal mirror. The infrared
images were rebinned by a factor of two to $0^{\prime\prime}\!.274$
pixel$^{-1}$ to better match to the seeing, which was $\approx
0^{\prime\prime}\!.9$.  Optical images were bias subtracted and flat
fielded using {\tt CCDPROC} in the {\tt IRAF} analysis package.
Infrared images were reduced using an in-house IRAF script which flat
fields, subtracts scaled sky images, shifts the images to a reference
image, then combines all images by averaging them.  The infrared sky
flat was created by averaging three bright sky images and three faint
sky images then taking the difference.  The sky flats of a particular
filter are taken every third night, where possible.  The flats from
one epoch to another do not differ by more than 1\%.

The final reduced images (Figure 7) were aligned to the USNO
astrometric reference frame.  Within the \chandra\ error circle, no
new source is seen in these optical or infrared images down to the
following limiting magnitudes: $V = 22.5, I = 21.3, J = 18.9, K =
17.5$.  These limits are referenced to the nearby star USNO B1.0
ID\#0702-0541769 whose magnitudes were given above.  The ($R-I$) and
($J-K$) colors of this object correspond to an M4 star.  Since there
is no $V$-band calibration for this field, the $V$ magnitude of the
reference star was therefore estimated from the ($V-I$) color of an M4
star. 

\section{Distance, Luminosity, and Timing Constraints on Interpretations}

The X-ray measured $N_{\rm H} = 1 \times 10^{22}$~cm$^{-2}$ is
somewhat more than half the total Galactic 21~cm measured $N_{\rm H} =
1.8 \times 10^{22}$~cm$^{-2}$ in this direction ($\ell =
10.\!^{\circ}73,\, b = -0.\!^{\circ}16$), which suggests only that the
distance to \src\ is of order 10~kpc.  However, it could be as close
as $3-5$~kpc if we adopt the typical run of visual extinction in the
solar neighborhood, $A_V = 1.5-2.0$ mag kpc$^{-1}$, together with
$N_{\rm H}/A_V = 1.6 \times 10^{21}$~cm$^{-2}$~mag$^{-1}$ (we note
that these estimates neglect possible additional systematic
uncertainty in $N_{\rm H}$ due to our choice of a particular X-ray
spectral model for fitting.)  More specifically, it is also useful to
compare with the properties of the supernova remnant G11.2--0.3, which
is only $0.\!^{\circ}5$ from \taxp.  An H~I absorption kinematic
distance of 5~kpc is estimated for G11.2--0.3 (Green et al. 1988).
Since the X-ray measured column density to G11.2--0.3 is $\approx 1.4
\times 10^{22}$~cm$^{-2}$ (Vasisht et al. 1996), and since the total
21~cm H~I column densities do not differ significantly between the two
X-ray source positions (Stark et al. 1992), we may consider that 5~kpc
is actually an upper limit on the distance to \taxp.  With the
distance somewhat uncertain, we parameterize the calculations that
follow in terms of $d_{5}$, the distance in units of 5~kpc.

The absence of an optical/IR counterpart for \src\ practically rules
out a high-mass binary companion, such as an OB supergiant or Be
transient.  For example, assume absolute magnitudes $M_V \approx -3.5$
and $M_K \approx -2.7$ as appropriate for stars in the spectral range
B0~V--B2~III.  Even in the presence of 6 magnitudes of $V$-band
extinction (the equivalent of $N_{\rm H} = 1 \times
10^{22}$~cm$^{-2}$), the $K$-band extinction is only 0.8 mag.  At a
distance of 5~kpc, a high-mass binary should therefore have $V < 16.0$
and $K < 11.6$, while the X-ray error circle is blank to $V > 22.5$
and $K > 17.5$ in the ANDICAM images.  Even if the column density were
underestimated by a factor of 5, the $K$-band extinction would be only
$\approx 3$~mag, not enough to obscure an OB companion in that band.

A low-mass binary, similar to the 7.67~s pulsar 4U~1626--67, is much
more difficult to exclude at this low Galactic latitude since a
late-type main-sequence or degenerate companion may fall below the
limits of our optical and IR images if its distance is toward the high
end of our estimates.  Consider a typical K5~V LMXB secondary of $M_V
\approx 7.3$ and $M_I \approx 6.9$.  At 5~kpc, its apparent magnitudes
including the effects of extinction could be $V = 26.8$ and $I =
23.3$.  Also, we may not necessarily expect to see optical emission
greatly enhanced by X-ray heating of a companion or outer accretion
disk, since the intrinsic X-ray luminosity in the 0.5--10~keV band is
only $3.9 \times 10^{35}\,d_{5}^2$ ergs~s$^{-1}$. 

The expected Doppler delay in a small LMXB orbit may also not be
detectable in the 5.54~s X-ray pulsations, a limitation that has long
thwarted timing tests for binary companions in AXPs in general
(Mereghetti, Israel, \& Stella 1998; Wilson et al. 1999).  We searched
for phase jitter in the XMM pulse profiles of \src\ by
cross-correlating folded light curves in 5~minute segments with a
master profile constructed from the entire observation.  No systematic
deviations from a constant phase were found to a limit of 0.005
cycles, which implies that $a_x\,{\rm sin}\,i$ of the neutron star's
hypothetical orbit must be less than 0.03 lt-s for orbital periods in
the restricted range 10~min -- 8~hr.  This is comparable to the limits
achieved in the most sensitive case for 1E~2259+586 by Mereghetti et
al. (1998).  As those authors discussed, such a limit rules out most
but not all main-sequence companions, but a helium-burning companion
or a white dwarf is still allowed.  Similarly, the known binary
companion of 4U~1626--67 is still not detectable via this method.  A
more restrictive search for a low-mass binary companion (or fossil
accretion disk) requires deeper optical/IR imaging.  Most recently,
the initial report by Israel et al. (2003) of a $K_s = 20.8$ object 
within the HRC error circle severely constrains the existence of such
counterparts.

The timing and spectral parameters of \taxp\ are typical of an AXP and
nearly identical to those of AXP 1E1048$-$59 (Paul et al. 2000), which
notably also lacks a detected SNR. The rapid spin-down rate ($\dot P =
[1.1-2.1] \times 10^{11}$ s/s; Ibrahim et al. 2003) and slow period
imply a large magnetic field ($B = 3.2 \times 10^{19} \sqrt{P \dot P}
\approx 3 \times 10^{14}$ G) and spin-down energy loss rate of $\dot E
\approx 4 \times 10^{33}$ erg~s$^{-1}$, far below the observed X-ray
luminosity, which is $3.9 \times 10^{35}\,d_{5}^2$ ergs~s$^{-1}$ in
the 0.5--10~keV band.  In the power-law plus blackbody model (Table~3)
the blackbody has a temperature of $7.7 \times 10^6$~K and a
bolometric luminosity of $1.2 \times 10^{35}\,d_{5}^2$ ergs~s$^{-1}$.
Its area is $\approx 6.0 \times 10^{11}\,d_{5}^2$~cm$^{2}$ (neglecting
unknown geometric and beaming factors), which is $\approx 5\%$ of the
area of a neutron star.  Thus, the large pulsed fraction of this
dominant component is consistent with rotational modulation of a
necessarily small emitting region on the surface.  This implies a flux
ratio of the two components of $F^{\rm pl}_X(2-10 \ \rm{keV})/F^{\rm
bb}_{\rm bol} = 0.8$.  These X-ray spectral properties are consistent
with the dependences on spin-down rate for AXPs and SGRs found by
Marsden \& White (2001), and also with the increase of pulsed fraction
with increasing ratio of $F^{\rm bb}_{\rm bol}/F_{\rm total}$ shown by
Israel, Mereghetti, \& Stella (2002).

In the alternative double blackbody spectral model (Table~3), the
fitted $N_{\rm H}$ is reduced to $6.3 \times 10^{21}$~cm$^{-2}$,
suggesting a smaller distance $d < 5$~kpc. The large surface area
implied for the cooler ($T = 3.4 \times 10^6$) blackbody component in
this model, $\approx 7.0 \times 10^{12}\,d_{5}^2$~cm$^{2}$ is $\sim$
half the area of the NS. This makes it difficult to explain why the
soft X-rays, which come from the cooler blackbody, have a pulsed
fraction as high as 36\% unless $d$ is significantly less than 5~kpc,
thus reducing the emitting area. In its favor, a purely thermal model
can explain why the pulsed fraction increases with energy, and why the
pulses are in phase at all energies, assuming the geometry to be that
of a small hot spot surrounded by a cooler annulus. Otherwise, the
observed pulsed fractions and phase relationship have no obvious
explanation in the power-law plus blackbody model.  So it is possible
that the X-ray manifestation of the ``outburst'' of a transient AXP is
largely evidence of a thermal heating event on the NS
surface. Moreover, we note that the temperature measured by \rosat\
using a single blackbody model is even lower that found for the
cooler component of the two-temperature model. When $N_{\rm H}$ is
held fixed at $6.3 \times 10^{21}$~cm$^{-2}$, the effective area of the
\rosat\ blackbody fit is $1.2 \times 10^{13} \ d^2_5$ cm$^2$,
compatible with the surface area of a NS. This perhaps accounts for the
failure to detect rotational modulation in the quiescent state.

\section{Conclusions and Suggestions for Further Work}

Although \src\ fell far from the location reported for \taxp\ by
Markwardt et al. (2003a), the detected pulsations clearly identify
them as the same source. Given the positional coincidence of the
\chandra\ and \xmm\ sources with the fainter \einstein, \rosat, and
\asca\ detections, we can only assume that they are one and the same
and summarize that the flux from \taxp\ has increased by nearly two
orders-of-magnitude between the year 2003 observations and those
fortuitously preserved in the archives. These archival observations
show suggest flux stability over the 9 years prior to the current
active state.  Since its first {\it pulsed} detection in 2003 January,
\taxp\ has faded by half, before which it was either too faint for
\xte\ to detect, or perhaps unpulsed.  The complete range of behavior
of \taxp\ has yet to be determined.

Our search for an optical/IR counterpart of \src\ is predicated on the
reliability of the \chandra\ error circle. Without a fiducial X-ray
source to register the HRC X-ray field against the optical reference
frame, we cannot eliminate the systematic portion of the aspect error
and have assumed the nominal uncertainty. Still, we can rule out any
bright optical source as a counterpart considering that the distance
to the closest star is $4\farcs7$, well in excess of the expected
systematic and random error.

Time variability of the flux from an isolated neutron star is
intriguing.  The flux change seen from \taxp\ is consistent with that
found for \axp, the original example of a transient AXP, which was
caught only once in a bright, active state.  The CCO in SNR RCW~103, a
non-pulsating AXP-like object (Gotthelf, Petre, \& Hwang 1997;
Gotthelf, Petre, \& Vasisht 1999), has been monitored for over a
decade and it is found to display marked variability on months to
years time scales.  Furthermore, like the CCO's, \axp\ is also
centered on a shell-type radio and X-ray SNR, G26.6+0.1 (Gaensler et
al. 1999, Vasisht et al. 2000).  If \taxp\ is an object related to
\axp\ and the CCO in RCW~103 as suggested by its spectrum and
variability, then we might expect to find an associated supernova
remnant, however, so far none is found either in X-ray images or in
the archival VLA NVSS map. This is most surprising as \taxp\ is one of
the younger AXPs, with a spin-down age ($P/2\dot P) \leq 7600$~yr,
based on the reported period derivative (Ibrahim et al. 2003).

The leading theory for the nature of SGRs and AXPs is the magnetar
model as first proposed by Duncan \& Thompson (1992).  In this model,
in the absence of soft-gamma-repeater-like outbursts, one expects
generally smooth spin-down, as found for \taxp\ (Ibrahim et al. 2003).
This model is also well suited to the properties of \axp, and is
consistent with the inferred magnetic field for \taxp\ (Ibrahim et
al. 2003).  However, while the magnetar theory as currently envisioned
explains the episodic hard X-ray activity associated with SGRs, it
does not predict the softer X-ray variability and pulsar turn on/off
seen in the TAXPs.  This behavior is more typically of accreting
binary systems or, perhaps hypothetically, accretion from a fall-back
disk of material that formed shortly after the supernova explosion
that gave birth to an isolated neutron star.  Severe constraints have
been placed on the plausibility of the latter scenario by optical/IR
observations of AXPs (Hulleman et al. 2000, Hulleman, van Kerkwijk, \&
Kulkarni 2000, Kern \& Martin 2002). In particular, direct detection
of high-amplitude optical or IR pulsations at the X-ray period, e.g.,
as Kern \& Martin (2002) achieved in the case of 4U0142+61, would be
strong evidence that \src\ is an isolated neutron star.

The discovery of a second example of a transient AXP, one which lacks
pulsation in its quiescent state, offers the possibility of
interpreting CCOs as quiescent AXPs.  No periodic signals have been
detected so far from the CCOs in the Cas~A and RCW~103 SNRs, for
example, despite deep timing observations of both. Cas~A is known to
be a very young ($\sim 300$ yrs) object with a spectrum consistent
with an AXP (Chakrabarty 2001; Mereghetti, Tiango, \& Israel 2002)
while SGRs are thought to be older ($\sim 10^4$ yrs) manifestations of
the AXPs (Gotthelf 1999; Gaensler 2001). AXPs and SGRs have long been
considered related phenomena, reinforced by the recent detection of
SGR-like bursts from two AXPs (Gavriil et al. 2002; Kaspi et
al. 2003). We hypothesize that the various ``classes'' of young
neutron stars that differ significantly from rotation-powered pulsars
are phenomenologically related, possibly through an evolutionary
progression.

Further monitoring of \taxp\ while it is in an active, pulsating state
is essential to determine the emission mechanism(s) and the time spent
at various levels of luminosity.  Monitoring can help determine the
probability of a neutron star being in a ``hidden'' inactive state of
high duty cycle. This has important consequences for population
studies of young neutron stars and further argument for a vast
underestimation of the AXP/NS formation rate (Gotthelf \& Vasisht
2000).  Measurement of the quiescent spectrum is especially important
to help identify this faint population of likely missed NSs.

\acknowledgements

We thank Fred Seward and Harvey Tananbaum for making the HRC
observations possible through the \chandra\ Director's Discretionary
Time program. We thank Craig Markwardt for providing updated error
contours for planning this observation. We especially thank Michael
Juda for advice on HRC issues and, along with Joy Nicole, expediting
the data processing and delivery.  We thank Fred Jansen for providing
the follow-up \xmm\ observation of \taxp\ and to Matthias Ehle and Leo
Metcalfe for their assistance in guiding the data delivery from a
difficult observation. Kudos to Maurice Leutenegger and Marc Audard
for their expert assistance with the \xmm\ data processing at Columbia
University. We are indebted to Andrew Rasmussen for making available
his preliminary RGS study of \taxp.  This research has made use of the
National Aeronautics and Space Administration (NASA)/IPAC Infrared
Science Archive, which is operated by the Jet Propulsion Laboratory,
California Institute of Technology, under contract with NASA. This
research has also made use of data obtained from the High Energy
Astrophysics Science Archive Research Center (HEASARC), provided by
NASA's Goddard Space Flight Center. This research is supported by NASA
LTSA grant NAG~5-8063 to E.V.G. C.B. and M.B. are supported by NSF
grant AST-0098421.

\clearpage

\begin{deluxetable}{lccccc}
\tabletypesize{\small}
\tablewidth{0pt}
\tablecaption{Imaging X-ray Observations of \taxp\ }
\tablehead{
\colhead{Mission/} & \colhead{Obs}  & \colhead{Exposure} & \colhead{R.A.}  & \colhead{Decl.} & \colhead{Count}   \\
\colhead{Instrument}& \colhead{Date} & \colhead{Time}     & \colhead{}      & \colhead{}      & \colhead{Rate\tablenotemark{a}} \\
\colhead{}        & \colhead{(UT)}  & \colhead{(ks)}      &\colhead{(J2000)}&\colhead{(J2000)}& \colhead{(s$^{-1})$}
}
\startdata
\einstein /IPC& 1980 Mar 31 & \s\ 0.7  & 18:09:53\ssss\  & --19:44:40\s\s\        & $0.03$   \\
\rosat /RASS  & 1990 Sep 03 & \s\ 0.3  & 18:09:51.5\s\      & --19:43:45.5\s         & $0.05$   \\
\rosat /PSPC  & 1991 Mar 18 & \s\ 3.0  & 18:09:49\ssss\  & --19:44\s\s\s\s\s\s\   & $0.03$   \\
\rosat /PSPC  & 1992 Mar 07 & \s\ 8.3  & 18:09:51.8\s\      & --19:43:35\s\s\        & $0.05$   \\
\rosat /PSPC  & 1993 Apr 02 & \s\ 10.0 & 18:09:49\ssss\  & --19:44\s\s\s\s\s\s\s\ & $0.04$   \\
\rosat /PSPC  & 1993 Apr 03 & \s\ 5.3  & 18:09:51.5\s\      & --19:43:45\s\s\        & $0.04$   \\
\asca /GIS    & 1996 Apr 02 & 11.0     & 18:09:51.3\s\      & --19:43:06\s\s\        & $0.02$   \\
\asca /GIS    & 1996 Apr 08 & 10.3     & 18:09:51.3\s\      & --19:43:06\s\s\        & $0.08$   \\
\asca /GIS    & 1999 Sep 28 & 38.8     & 18:09:51.3\s\      & --19:43:06\s\s\        & $0.08$   \\
\chandra /HRC & 2003 Aug 27 & \s\ 2.8  & 18:09:51.13        & --19:43:51.7           & $0.96$   \\
\xmm /EPIC-pn & 2003 Sep 08 & \s\ 11.5 & 18:09:51.03        & --19:43:51.1           & $10.5$   \\
\chandra /HRC & 2003 Nov 01 & \s\ 3.0  & 18:09:51.08        & --19:43:51.7           & $1.13$   \\
\enddata
\tablenotetext{a}{\footnotesize Background subtracted count rate corrected for detector dead-time. Rates quoted for \asca\ are for the combined GIS2+GIS3 detectors.}
\label{ta:log}
\end{deluxetable}


\begin{deluxetable}{lcccccc}
\tablewidth{0pt}
\tablecaption{Timing Results}
\tablehead{
\colhead{Mission/}    & \colhead{Obs Date} & \colhead{Epoch\tablenotemark{a}}     & \colhead{Period} \\
\colhead{Instrument}  & \colhead{(UT)}     & \colhead{(MJD/TDB)} & \colhead{(s)}
}
\startdata
\chandra /HRC & 2003 Aug 27  & 52878.96851749   &  $5.5392 \pm 0.0008$   \\
\xmm /EPIC-pn & 2003 Sep 08  & 52890.56420438   &     $5.539344 \pm 0.000019$ \\
\chandra /HRC & 2003 Nov 01  & 52944.62890753   &  $5.5391 \pm 0.0006$   \\
\enddata
\tablenotetext{a}{\footnotesize Epoch of phase zero in Figures 3 and 4. Error are quoted at the 95\% confidence level}
\label{ta:timing}
\end{deluxetable}


\begin{deluxetable}{lcccccc}
\tabletypesize{\footnotesize}
\tablewidth{0pt}
\tablecaption{Spectral Fits and Fluxes}
\footnotesize
\tablehead{
\colhead{Mission/ }      & \colhead{Obs Date} & \colhead{$N_{\rm_H}$}           & \colhead{$\Gamma$ or $kT$} & \colhead{$kT$} & \colhead{Flux\tablenotemark{a}} & \colhead{$\chi^2_{\nu}$(DoF)} \\
\colhead{Instrument}          & \colhead{(UT)}& \colhead{($10^{22}$ cm$^{-2}$)}    & \colhead{(keV)}         &\colhead{(keV)} & \colhead{(ergs cm$^{-2}$ s$^{-1}$)}&  
}
\startdata
\einstein /IPC & 1980 Mar 31 &$0.63$(fixed)   & \dots          & $0.18$(fixed)           & \ss\ $10 \times 10^{-13}$  & \dots \\
\rosat /RASS  & 1990 Sep 03 & $0.63$(fixed)   & \dots          & $0.18$(fixed)           & \s\s\s\s\s\ $5 \times 10^{-13}$ & \dots \\
\rosat /PSPC  & 1991 Mar 18 & $0.63$(fixed)   & \dots          & $0.18$(fixed)           & \s\s\s\s\s\ $7 \times 10^{-13}$ & \dots \\
              & 1992 Mar 07 & $0.63$(fixed)   & \dots          & $0.18 \pm 0.02$         & \s\s\s\ $6.9 \times 10^{-13}$ & 1.1(13) \\
              & 1993 Apr 02 & $0.63$(fixed)   & \dots          & $0.19 \pm 0.02$         & \s\s\s\ $8.3 \times 10^{-13}$ & 1.7(14) \\
              & 1993 Apr 03 & $0.63$(fixed)   & \dots          & $0.18 \pm 0.02$         & \s\ $5.5 \times 10^{-13}$  & 0.8(14)   \\
\asca /GIS    & 1996 Apr 02 & $0.63$(fixed)   & \dots          & $0.22 \pm 0.07$         & \s\ $8.1 \times 10^{-13}$  & 0.5(10)  \\
              & 1996 Apr 08 & $0.63$(fixed)   & \dots          & $0.18$(fixed)           & \s\ $7.5 \times 10^{-13}$  & \dots    \\
              & 1999 Sep 28 & $0.63$(fixed)   & \dots          & $0.19 \pm 0.04$         & \s\ $6.5 \times 10^{-13}$  & 0.4(20)  \\
\hline
\chandra /HRC & 2003 Aug 27 & $1.05$(fixed)   & $3.7$ (fixed)  & $0.67$ (fixed)          & $3.92 \times 10^{-11}$ & \dots    \\
                 &             & $0.63$(fixed)   & $0.29$ (fixed) & $0.70$ (fixed)          & $3.70 \times 10^{-11}$ & \dots    \\
\hline
\xmm /        & 2003 Sep 08 & $1.05 \pm 0.05$ & $3.7 \pm 0.2$  & $0.67 \pm 0.01$         & $3.98 \times 10^{-11}$ & 1.0(193) \\
EPIC-pn       &             &                 &                & BB flux:         & $2.63 \times 10^{-11}$ &          \\
PL+BB model   &             &                 &                & PL flux:         & $1.35 \times 10^{-11}$ &          \\
\hline
\xmm /        & 2003 Sep 08 & $0.63 \pm 0.05$ & $0.29 \pm 0.03$& $0.70 \pm 0.02$         & $3.94 \times 10^{-11}$ & 1.0(190) \\
EPIC-pn       &             &                 &                & BB1 flux:        & $5.36 \times 10^{-12}$ &          \\
BB1+BB2 model &             &                 &                & BB2 flux:        & $3.40 \times 10^{-11}$ &          \\
\hline
\chandra /HRC & 2003 Nov 01  & $1.05$ (fixed)  & \s\s\ $3.7$ (fixed)  & $0.67$ (fixed)          & $3.35 \times 10^{-11}$ & \dots    \\
              &              & $0.63$ (fixed)  & $0.29$ (fixed) & $0.70$ (fixed)          & $3.17 \times 10^{-11}$ & \dots    \\
\enddata
\tablenotetext{a}{\footnotesize Absorbed flux in the $0.5-10$ keV band. All errors are 90\% confidence level for a single interesting parameters.}
\label{ta:spectra}
\end{deluxetable}

\clearpage

\begin{figure}
\begin{minipage}{0.45\linewidth}
\centerline{
\plotone{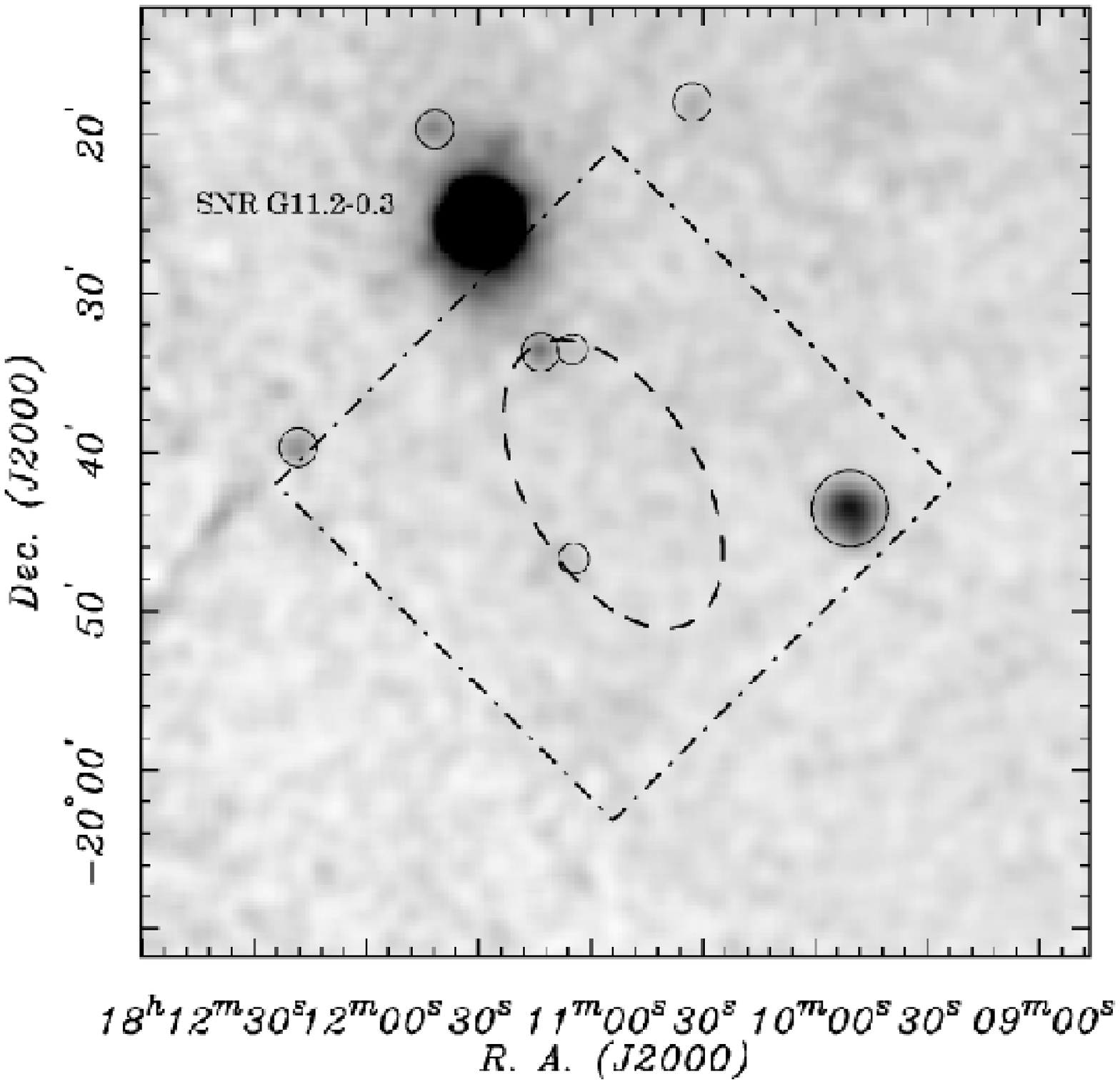}
}
\caption{The original \xte\ 99\% confidence ellipse ({\it dashed
line}) for \taxp\ shown overlaid on the \rosat\ PSPC image of the
local environs. The image is saturated to highlight faint  \rosat\ and 
\asca\ sources, indicated by the solid circles.
The \chandra\ HRC-I field of view is outlined by the {\it dot-dashed\/} square.
}
\end{minipage}
\begin{minipage}{0.45\linewidth}
\centerline{
\plotone{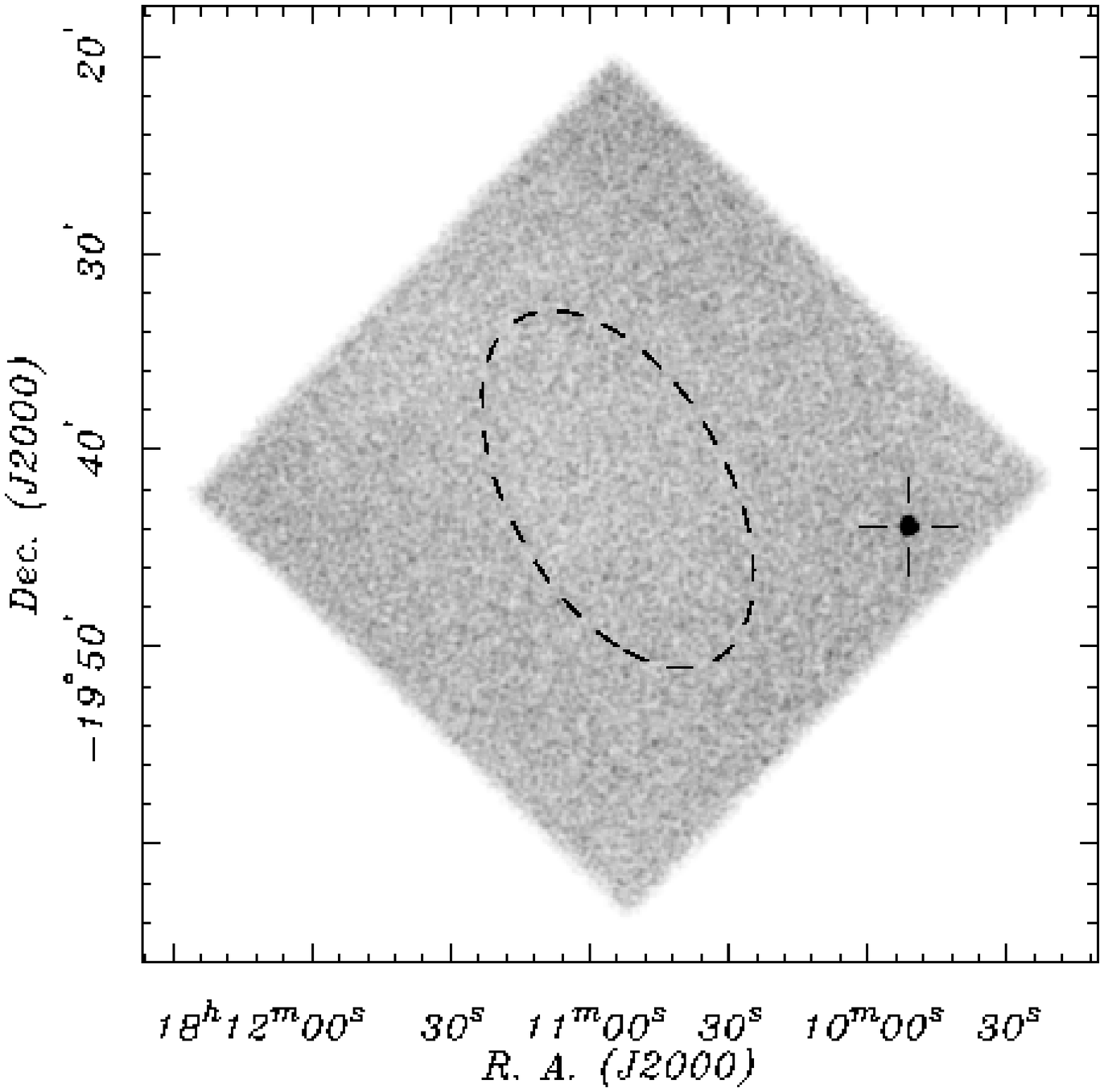}
}
\caption{ The intial \chandra\ HRC-I image of the \taxp\ field.  The {\it
dashed line} indicates the original \xte\ 99\% confidence ellipse for
\taxp.  Only one source, \src, is detected, which turns out to be
\taxp.  Its position is consistent with that of a \rosat\ source seen in
Figure~1.}
\end{minipage}
\end{figure}

\begin{figure}
\begin{center}
\begin{minipage}{0.45\linewidth}
\plotone{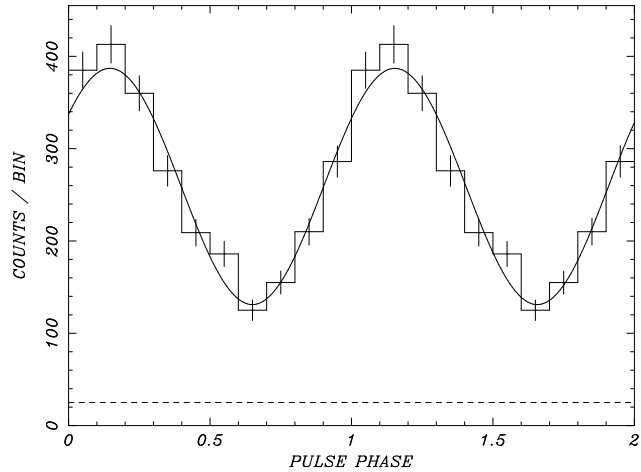}

\caption{ The  2003 Aug 27 \chandra\ HRC-I folded light curve of \src, identified
with the 5.54~s transient pulsar \taxp. The best fit sinusoidal model
({\it solid line}) and background level {\it dashed line} are
indicated. The pulsed fraction (defined in the text) for the signal is
$56\pm 7$\%.  The epoch of phase zero is given in Table~2.  Two cycles are
shown for clarity. }
\end{minipage}
\end{center}
\end{figure}

\clearpage

\begin{figure}
\centerline{
\epsscale{0.85}
\plotone{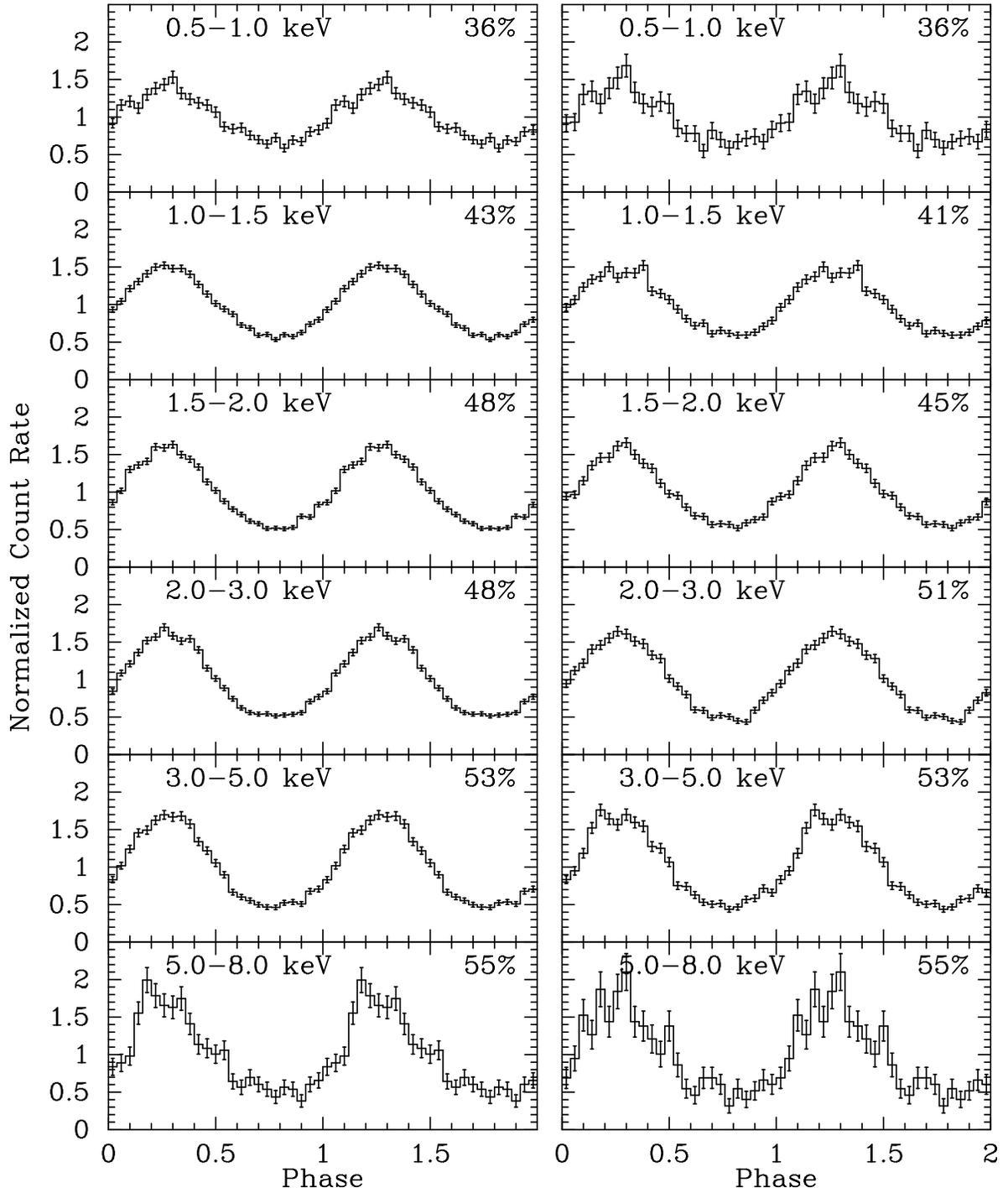}
}
\caption{ Pulse profiles of \src\ from the EPIC pn ({\it left}) and MOS1
({\it right}) CCDs, folded at the 5.539344~s period.  The epoch of phase
zero is given in Table~2.  Pulsed fractions are indicated in each panel.
Background has been subtracted for the EPIC~pn, but
not for the MOS because of the small size of its window. However,
the background is negligible in almost all bands.}
\end{figure}

\clearpage

\begin{figure}
\centerline{
\epsscale{0.6}
\plotone{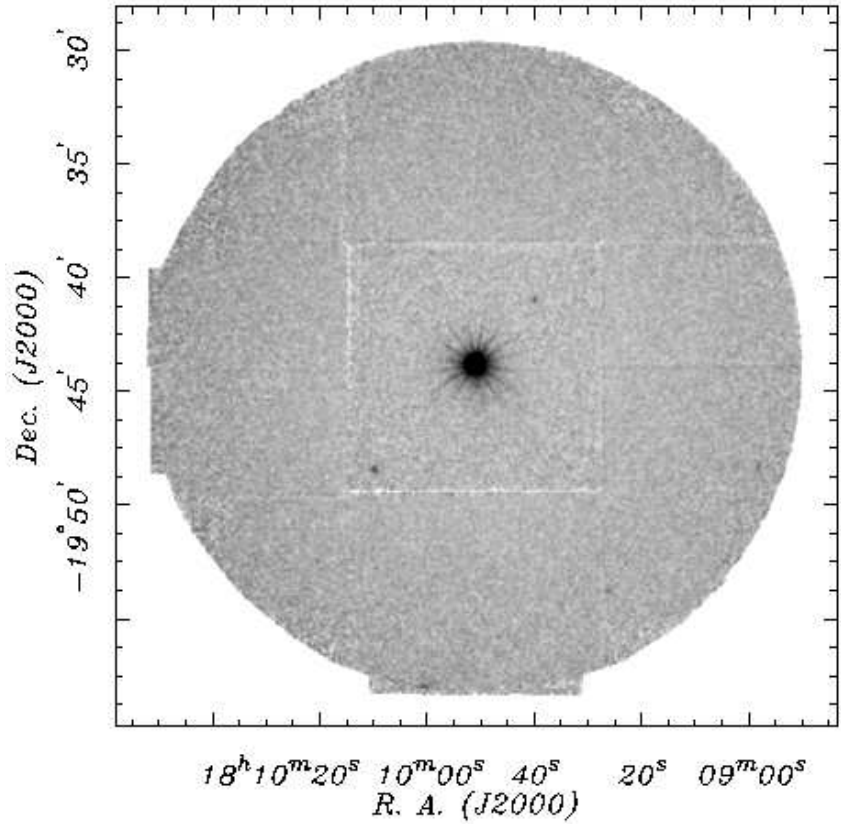}
}
\caption{ The \xmm\ EPIC-MOS image of \taxp\ herein identified with \src\ (central source).
No evidence is seen for an associated X-ray supernova remnant.
}
\end{figure}

\bigskip

\begin{figure}
\centerline{
\epsscale{0.6}
\plotone{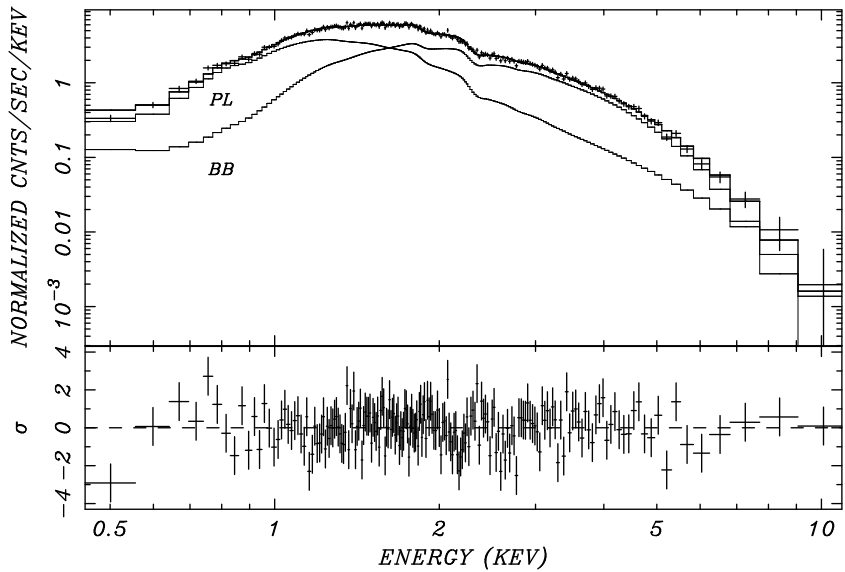}
}
\caption{ The \xmm\ EPIC-pn spectrum of \taxp\ fitted with the
two-component blackbody (BB) plus power-law (PL) model described in
the text. The contribution of each spectral component is indicated by
the solid lines. The lower panel shows the residuals from the best fit
model. No evidence is found for narrow lines or cyclotron absorption
features.}
\end{figure}

\bigskip

\begin{figure}
\centerline{
\epsscale{0.5}
\plotone{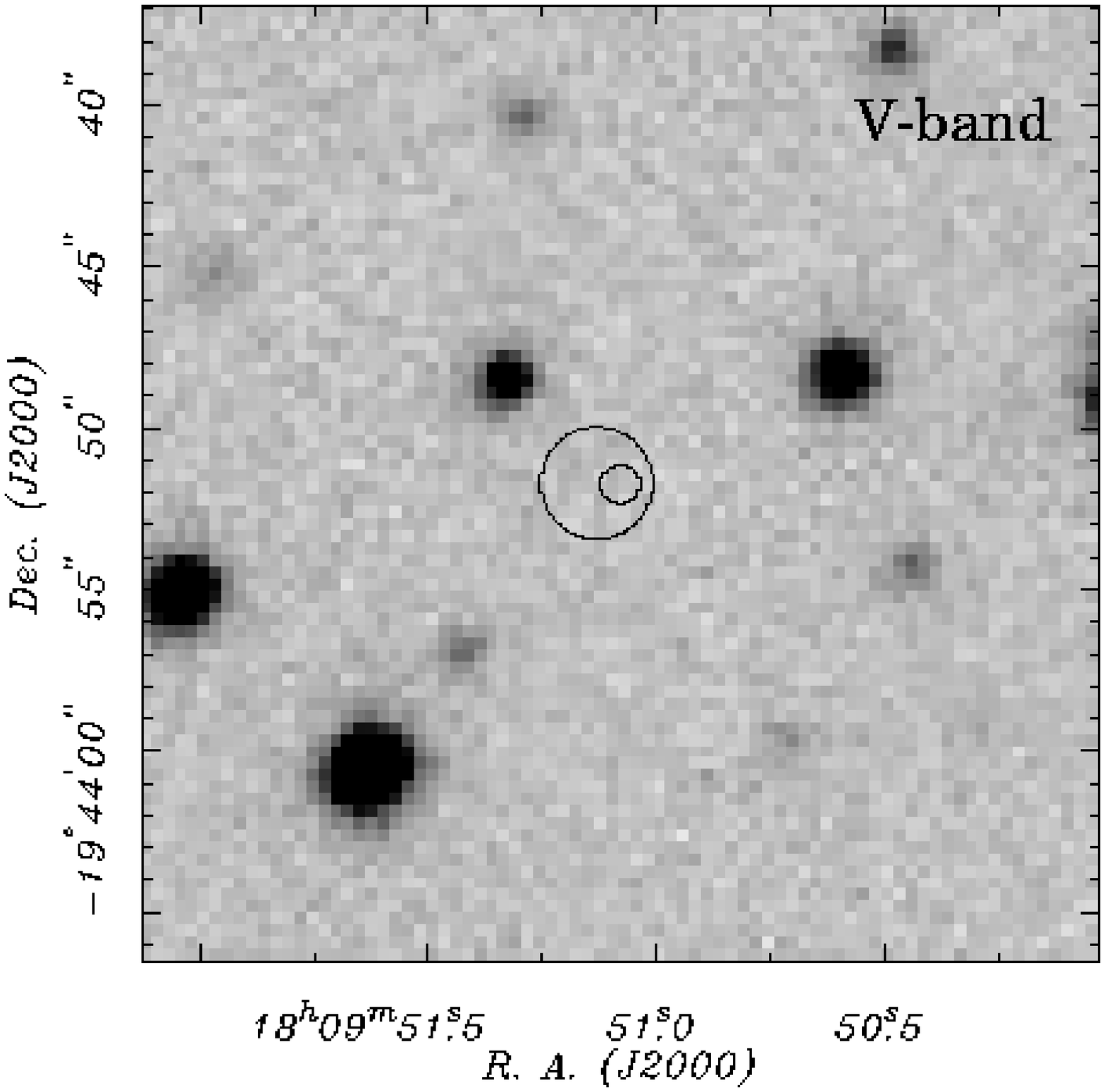}
\hfill
\epsscale{0.5}
\plotone{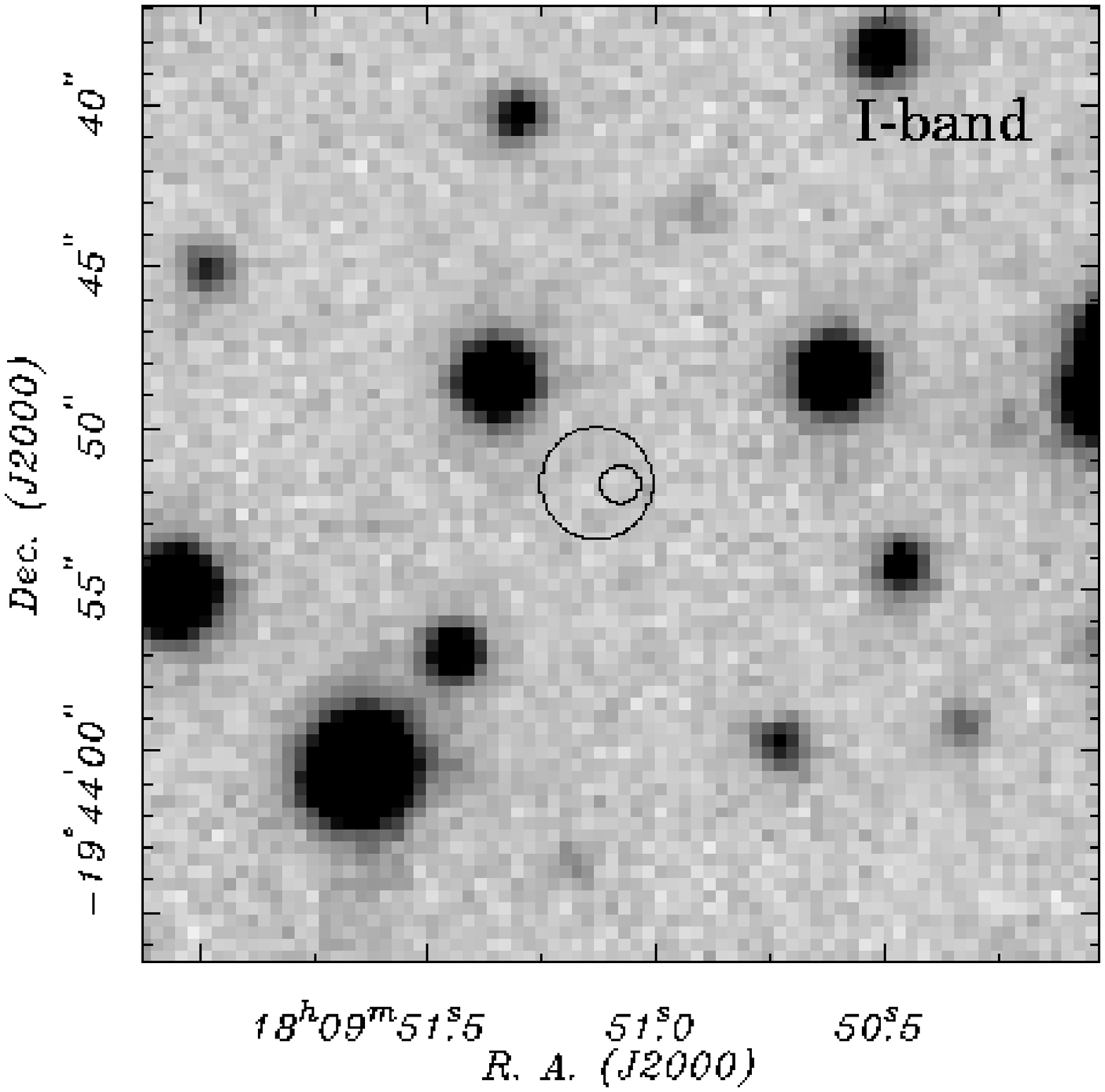}
}
\bigskip
\centerline{
\epsscale{0.5}
\plotone{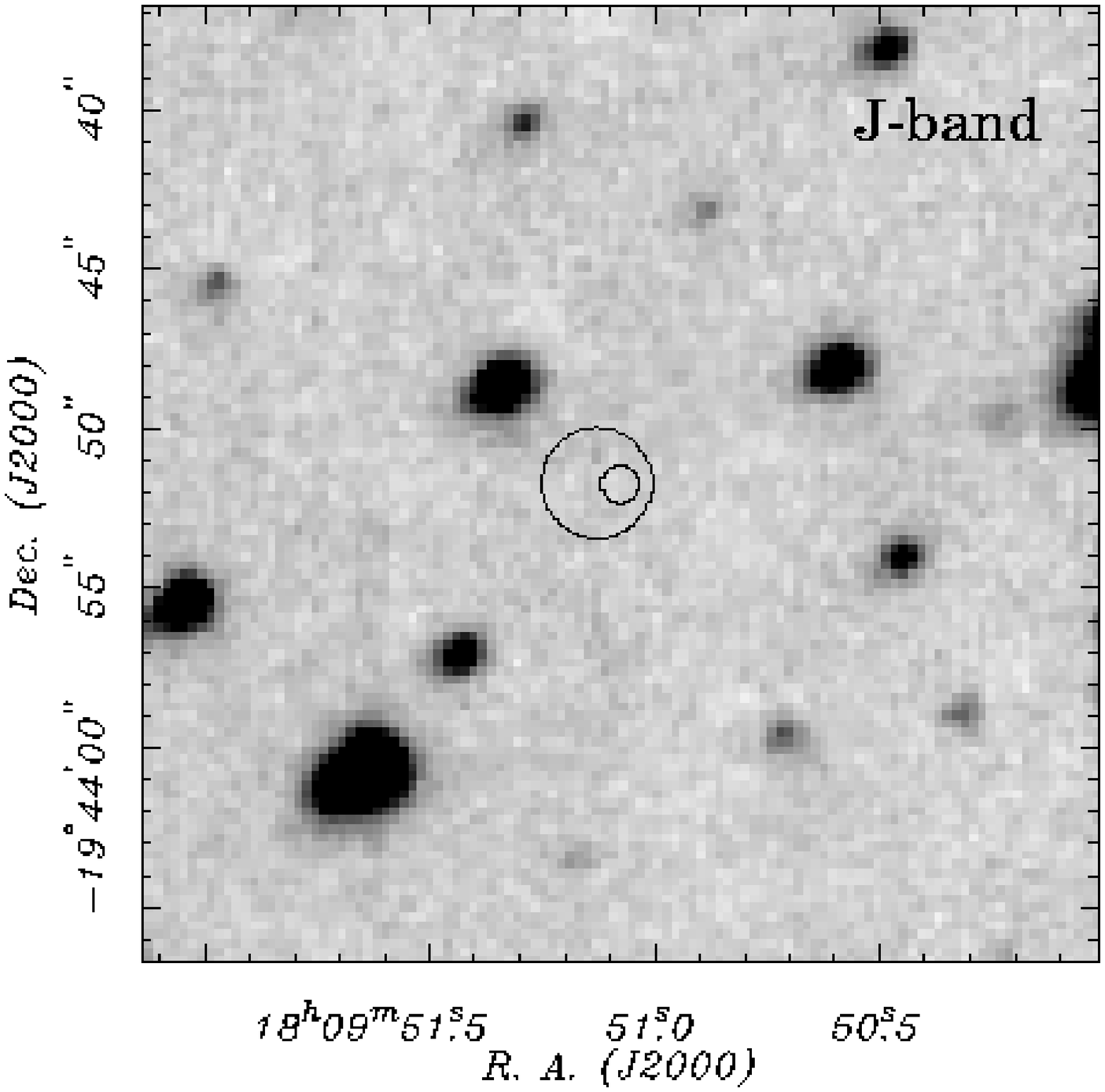}
\hfill
\epsscale{0.5}
\plotone{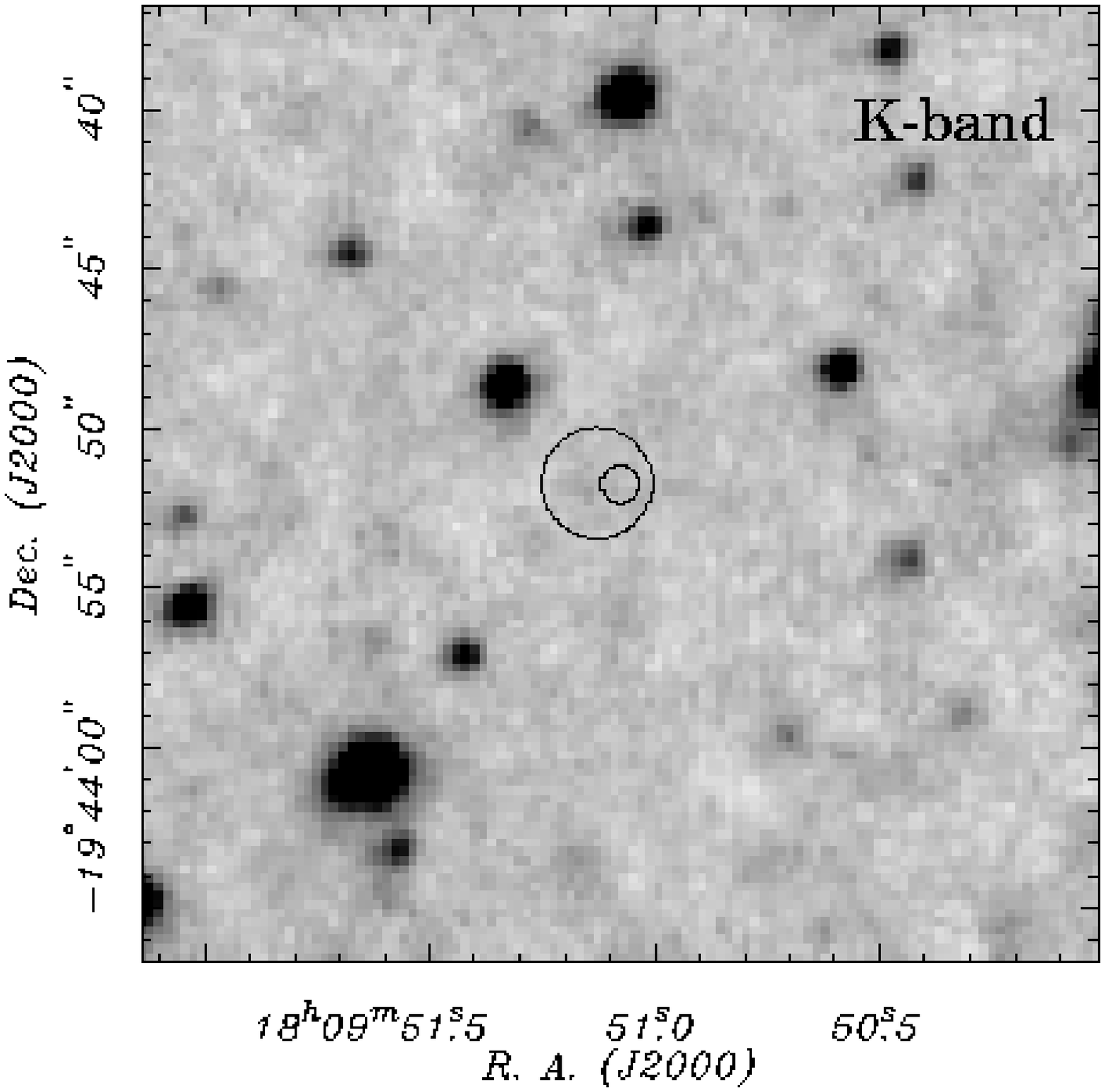}
}
\caption{Optical and infrared images of the field of \src, taken with
the ANDICAM instrument on the SMARTS 1.3~m telescope at CTIO.  The
circles indicate the $1\farcs7$ and $0\farcs6$ radius (90\% confidence level)
\chandra\ HRC localization (see text). Limiting magnitudes of these
images are $V = 22.5, I = 21.3, J = 18.9, K = 17.5$.  }
\end{figure}

\end{document}